\documentclass[12pt]{article}
\usepackage{fullpage}

\usepackage{graphicx,psfrag,amsmath,url}
\newcommand{\BEAS}{\begin{eqnarray*}}
\newcommand{\EEAS}{\end{eqnarray*}}
\newcommand{\BEQ}{\begin{equation}}
\newcommand{\EEQ}{\end{equation}}
\newcommand{\BIT}{\begin{itemize}}
\newcommand{\EIT}{\end{itemize}}

\newcommand{\eg}{{\it e.g.}}
\newcommand{\ie}{{\it i.e.}}

\newcommand{\reals}{{\mbox{\bf R}}}

\newcommand{\argmin}{\mathop{\rm argmin}}

\newcounter{oursection}

\usepackage{array}
\usepackage{algorithm2e}
\usepackage{caption,subcaption}
\usepackage{multirow}
\usepackage{versions}

\excludeversion{derivations}

\includeversion{keywords}

\newif\ifarxiv
\arxivtrue

\usepackage{circuitikz}
\usepackage{tikz}

\title{Value Function Approximation for Direct Control of Switched Power Converters}

\author{
Nicholas Moehle\thanks{
Mechanical Engineering Department, Stanford University. 
\texttt{moehle@stanford.edu}}
\and Stephen Boyd\thanks{
Electrical Engineering Department, Stanford University. 
\texttt{boyd@stanford.edu}}
}
\date{\today}

\begin{document}

\maketitle

\begin{abstract}
We consider the problem of controlling switched-mode power converters
using model predictive control.  Model predictive control 
requires solving optimization problems in real time,
limiting its application to systems with small numbers of switches and a short horizon.  
We propose a technique for using off-line computation
to approximate the model predictive controller.
This is done by dividing the planning horizon into two segments,
and using a quadratic function to approximate the optimal cost over the second segment.
The approximate model predictive algorithm minimizes the true cost over the first segment,
and the approximate cost over the second segment,
allowing the user to adjust the computational requirements by changing the length
of the first segment.
We conclude with two simulated examples.
\end{abstract}


\section{Introduction}

\paragraph{Predictive control.}
Model predictive control (MPC) is emerging as a powerful tool for
controlling switched-mode power converters.
A model predictive control scheme uses a planning model
to predict the effects of control decisions
on the converter over some finite planning horizon.
It then compares the outcomes of these decisions using a cost criterion,
choosing an optimal control input sequence, \ie, one
that minimizes the cost over the horizon.
The controller then uses the first control input in the trajectory,
waits a specified time, then repeats the process.
Model predictive control is now widely used in many control applications,
particularly those that are sampled slowly enough that the required
minimization problem can be solved in the sample time.

Finite-control-set MPC (FCS-MPC)
refers to the case in which
the control input takes on a finite number of values.
This is particularly relevant for switched-mode power converters,
where the control inputs typically represent allowable switch configurations.
In this case, solving the optimization problem only requires evaluating
the objective function a finite number of times
and choosing the control inputs that achieve the lowest cost.
When the total number of control inputs over the planning horizon
is small,
this can be done quickly on modest computational platforms,
which is attractive for embedded control.


Unfortunately, the computational cost of solving the FCS-MPC problem
increases exponentially with the length of the planning horizon,
which means that checking all combinations
of control input trajectories becomes very difficult
for discrete horizon lengths of more than, say, four or five.
In practice, this means FCS-MPC is only useful
for converters with simple dynamics,
for which short planning horizons (and myopic controllers) suffice,
or converters with powerful processors available for control.
Even worse, many converter topologies have dynamics
that inherently require a long planning horizon for FCS-MPC to work well,
such as those with inverse response (\ie, non-minimum phase behavior).
For example, to increase the steady-state output voltage of an ideal boost converter,
it may be necessary to charge the inductor,
which can have the short-term effect of \emph{decreasing} the
output voltage.
A long planning horizon may be necessary
for the controller to predict the benefits of charging the inductor
before increasing the output voltage.
We illustrate this behavior for a boost converter in \S \ref{s-boost-example}.
(For further discussion on non-minimum phase behavior in power converters,
see \cite[Ch. 9]{erickson2007fundamentals} and \cite{karamanakos2014direct}.)
 
\paragraph{Our approach.}
In this paper, 
we provide a technique for \emph{approximately} solving
the optimization problems arising in FCS-MPC
with a long planning horizon, with very low computational cost.
This allows the resulting approximate FCS-MPC controller to make 
near-optimal decisions, with long prediction horizons,
at only a tiny fraction of the computational cost of
traditional long-horizon FCS-MPC.

To do this, we split the planning horizon into two segments.
The dynamics and cost function over the first segment are represented exactly,
and the problem is solved by exhaustive search,
as in traditional FCS-MPC.
The optimal cost over the second (typically much longer) 
segment is approximated using 
a multivariate quadratic function,
called the \emph{approximate value function}.
This function is determined during a
pre-solve phase (\ie, during controller synthesis),
using a desktop computer or server.
The controller itself simply compares a relatively
small number of possible input sequences with a short horizon,
and chooses the best one, taking into account the approximate value function.
Our method therefore shifts computational burden from
real-time execution to an off-line, precomputation phase,
which allows us to execute FCS-MPC-like controllers 
with very long horizon lengths on very modest computational hardware.

The resulting control algorithm, which we call 
\emph{value function approximation model predictive control}, 
which we shorten to \emph{approximate model predictive control} (A-MPC)
in this paper,
is very similar to short-horizon FCS-MPC,
differing only by the addition of an extra cost term 
(given by the approximate value function),
which represents the approximate long-term cost.
This resemblance is a practical and pedagogical advantage,
as our method can be explained and implemented as a simple modification
of short-horizon FCS-MPC
that imparts better dynamics on the closed-loop system,
since it has the advantage of a long planning horizon.

The price of doing this is decreased flexibility:
because deriving an approximate value function often 
requires extensive offline precomputation,
our formulation does not handle gross
changes in the objective function or dynamics,
although some moderate changes, such as changing the desired state value,
can be easily accommodated.
However, we suspect that such large-scale changes in the 
operating requirements are rare,
and in any case, can be handled by synthesizing several controllers,
and switching between them as desired.
(This is similar to gain scheduling linear controllers.)


\subsection{Related work}

\paragraph{FCS-MPC for power converters.}
A good introduction to FCS-MPC
is given by Kuoro \cite{kouro2009model}
and Cort\'{e}s \cite{cortes2008predictive},
which include many references to practical applications
as well as tricks that can be used with FCS-MPC
(\eg, shaping the switching spectrum
or controlling reactive power).
These papers primarily cover short-horizon FCS-MPC,
which is easily solved by brute-force enumeration of all possible switching trajectories.

For long-horizon FCS-MPC, enumerating all switching trajectories
is not computationally feasible,
and more sophisticated techniques are used to solve
(or approximately solve) the optimization problem.
For example, branch and bound uses lower bounds on the optimal cost 
to identify and remove suboptimal parts of
the decision tree (see \cite{boyd2007branch}).
Simple lower bounds for the FCS-MPC problem are proposed by Geyer 
\cite{geyer2011computationally},
by simply assuming zero future cost starting from any node in the decision tree.
Another general technique is move blocking
(\ie, artificially limiting the number of control input trajectories
to obtain a suboptimal but practical solution);
an application to boost converter control can be found in 
\cite{karamanakos2012direct}.
Another recent approach is to use a sphere decoding algorithm to solve
the FCS-MPC optimal control problem \cite{geyer2013multistep}. 
Additionally, techniques from hybrid system theory can be used to explicitly
write the solution of the FCS-MPC problem
as a lookup table with regions defined by linear or quadratic inqualities.
(see \cite{borrelli} for details).
Unfortunately, the same exponential blowup that occurs in long-horizon
FCS-MPC can appear in explicit solutions of hybrid MPC,
\ie, it is possible that there are an exponential number of regions
in the lookup table.
There have been many applications of this approach to power converters;
see \cite{quevedo2014predictive} and \cite{geyer2005model},
and references therein.

\paragraph{Other control techniques.}
Most traditional techniques for controlling power converters
are based on linear control.
Because power converters are switched nonlinear systems,
it is first necessary to linearize the system around an operating point,
and use PWM, or a similar modulation scheme, to convert the continuous
control decisions of the linear controller to discrete switching decisions.
For an introduction to these techniques, see 
\cite{kassakian1991principles} and \cite{erickson2007fundamentals}.
For a discussion of continuous MPC techniques (and others) applied to
controller power converters, see \cite{wang2014pid}.
Deadbeat control policies attempt to drive the system state to
a desired value as quickly as possible,
often using a short prediction horizon.
A good summary can be found in \cite{quevedo2014predictive};
applications to power converters can be found 
in the references of \cite{kouro2009model}.

Many control methods are based on explicitly defining the switching boundaries
in state space,
usually defined using first- or second-order equations;
this idea originated with Burns and Wilson
\cite{burns1976state}, 
\cite{burns1977analytic}.
For some simple converters,
these boundaries can be derived as 
solutions to certain optimal control problems 
using Pontryagin's maximum principle.
For example, the solution to the minimum-time state transfer problem
for a buck converter has a switching boundaries defined by second-order equations.
However, this approach becomes unwieldy for even slightly more complicated systems
(see \cite{dhople2014time} for a discussion).


\paragraph{Dynamic programming.}
Recovering optimal control policies through value function approximation,
which is the basis of the approach in this paper,
is the subject of approximate dynamic programming,
and has been studied extensively in 
operations research, control, robotics, and other fields
(see \cite{bertsekas1995dynamic} and \cite{powell2007approximate} for an introduction).
The general method of exactly solving an optimal control problem
over some time steps, and using an approximation of the value function
for the rest of the horizon, is well known in the dynamic programming
literature, with names such as roll-out policy or look-ahead policy.
Adaptive versions of these methods,
which change the approximate value function during operation of the system,
are possible;
algorithms of these types are the subject of reinforcement learning;
see \cite{sutton1998introduction} and \cite{bertsekas1995dynamic}.

Our algorithm relies on quadratic approximate value functions,
obtained by simple function fitting techniques;
other methods for obtaining quadratic approximate value functions for switched-affine
systems can be found in 
\cite{rantzer2006relaxed} and \cite{wang2015approximate}.
Fast evaluation of control policies involving quadratic value functions
(as well as much more general problems)
can be found in 
\cite{yang2011lyapunov}, \cite{mattingley2011receding}, and
\cite{yang2010fast}.

\paragraph{Mixed-integer nonlinear programming.}
The optimal control problem solved in an FCS-MPC control policy
is a special case of mixed-integer nonlinear programming (MINLP).
MINLP problems are difficult to solve in general,
but algorithms exist that appear to work well in practice.
Several special cases,
such as mixed integer linear programming \cite{wolsey1998integer},
and mixed integer second-order cone programming \cite{benson2013mixed},
have dedicated solvers
(see Gurobi \cite{gurobi}, Mosek \cite{andersen2000mosek}, and CPLEX),
which often outperform general MINLP solvers.
In many cases, small problems 
(such as the FCS-MPC problem for power converters)
can be solved on a desktop in a few seconds.
Furthermore, these solvers can often be interfaced through modeling languages,
such as CVXPY \cite{cvxpy}, and CVX \cite{cvx, gb08},
which increase accessibility and decrease development time.


\subsection{Contributions}
In this paper we present a tractable approximate method
for solving long-horizon predictive control (FCS-MPC)
problems for switched-mode power converters.
We provide a theoretical justification for our method,
as well as practical approach for computing this added extra cost term.
We give simulated results for controlling a boost converter control
and an inverter, 
and we show that our method performs markedly better short-horizon predictive control
and performs comparably to long-horizon predictive control,
at a tiny fraction of the computational cost.

We contrast our method with other methods to achieve long-horizon FCS-MPC.
Many methods attempt to solve the optimal control problem
on the fly, which often necessitates a fast control processor.
Our method, by contrast, 
can be viewed as a simple modification of short-horizon
(or even one-step) FCS-MPC.
This allows our method to work with very low computational cost,
so that our method can be executed
in microseconds or less on modest computational platforms,
while still achieving some of the benefits of long-horizon MPC methods.


\subsection{Outline}
In \S\ref{s-mpc}, we present the theoretical basis for our method,
starting with the mathematical optimal control problem
to be solved by the controller,
and our general approach for approximately solving it.
In \S\ref{s-function-fitting}
we show how to numerically derive the controller
by approximating the value cost as a quadratic function.
In \S\ref{s-implementation}, we show how to implement the controller,
including several special cases in which extra computational savings can be attained.
In \S\ref{s-examples}, we demonstrate our approach on 
two example converters, a boost converter and an inverter.
In \S\ref{s-conclucion}, we conclude with some ideas for extending our method,
and alternative techniques for computing the extra cost term.

\section{Model predictive control}
\label{s-mpc}

\paragraph{Switched dynamical systems.}
We consider the switched dynamical system
\begin{align}
\label{e-sw-sys}
x_{t+1} = f(x_t, u_t),
\quad
t = 0, 1, \dots,
\end{align}
with state $x_t \in \reals^n$ and 
(finite) control input $u_t \in \{1, \dots, K\}$.
The subscript $t$ denotes (sampled) time.

In a converter, the state typically consists of physical variables 
such as inductor currents and capacitor voltages, 
but it can also include other, non-physical variables such as
a running sum of errors.
The input corresponds to the allowed switch configurations, and $K$ is
the number of allowed switch positions.
The dynamics function $f$ gives the state at the next sample time,
given the current state and choice of switch configuration.

\paragraph{Control.}
Our task is to choose a \emph{control policy} 
$\phi:\reals^n \to  \{1, \dots, K\}$
that maps the system state $x_t$ at time $t$ to a control input $u_t$, so that
\[
u_t = \phi(x_t).
\]
We design the policy according to some objectives, 
such as maintaining the state near some desired value,
or minimizing the converter power loss.
The policy divides the state space into $K$ 
(possibly disconnected) regions,
so that if the state is in the $i$th region,
the controller chooses input $i$.
The boundaries of these surfaces are often known as switching surfaces
or switching boundaries;
when the state of the system crosses a switching surface, the 
input (\ie, switch configuration) changes.
Implicit here is that we can measure the state; when this is not the case,
the state can be estimated using, for example, a Kalman filter.

\paragraph{Model predictive control.}

Under a model predictive control policy $\phi_{\rm mpc}$,
the control input is chosen by solving an optimal control problem:
to find $\phi_{\rm mpc}(z)$,
we first solve the finite-horizon optimal control problem
\begin{equation}
\begin{array}{ll}
\mbox{minimize} & 
  \sum_{t=0}^{T - 1} g(\tilde x_t) + h(\tilde x_{T}) \\
\mbox{subject to} & \tilde x_{t+1} = f(\tilde x_t, \tilde u_t) \\
                  & \tilde x_0 = z.
\end{array}
\label{e-opt-ctrl}
\end{equation}
The decision variables are the predicted system states $\tilde x_{0}, \dots, \tilde x_{T}$
and the planned control inputs 
$\tilde u_0, \dots, \tilde u_{T-1}$.
The integer $T$ is the horizon; it determines how far ahead in time we plan
when we choose the current input.
The function $g: \reals^n \to \reals$ is the stage cost function,
and the function $h: \reals^n \to \reals$ is the terminal cost function.
Then $\phi_{\rm mpc}(z)$ is chosen to be $\tilde u_0$, where $\tilde u$ is
any solution of the optimal control problem~(\ref{e-opt-ctrl}).




To specify an MPC policy, the designer must choose the horizon $T$,
the stage cost function $g$, and the terminal cost $h$.
The main difficulty of implementing the MPC policy $\phi_{\rm mpc}$
is quickly solving (\ref{e-opt-ctrl}), 
which is (in general) a mixed-integer nonlinear program (MINLP).
In principle the problem can be solved by 
checking the objective value for each possible control input trajectory,
\ie, for each of the $K^T$ possible trajectories,
and simply simulating the dynamics for each choice of input sequence.
However, for large values of $K$ and $T$ (larger than, say $K =3$ and $T=5$),
this approach is not practical.
In this case, (\ref{e-opt-ctrl}) is very difficult to solve exactly 
on an embedded processor at sub-millisecond speeds,
required for real-time control of switched power converters.
We note, however, that many effective algorithms exist
to solve MINLP problems reasonably quickly 
(if not at sub-millisecond time scale) on desktop computers or servers,
which we will return to later.


A large value of the planning horizon $T$ allows a receding-horizon controller
to consider the impact of current actions far in the future,
and almost always results in a superior control law.
Using a short planning horizon results in a myopic controller,
which can seriously degrade performance. 
Unfortunately, our desire for long horizon directly contradicts
our desire to easily solve \eqref{e-opt-ctrl} using low-cost, embedded processors,
because a large value of $T$ makes \eqref{e-opt-ctrl} much more difficult to solve numerically.
This effectively limits receding-horizon controllers based on \eqref{e-opt-ctrl}
to converters with simple dynamics 
(for which myopic controllers perform well)
or converters with powerful processors available for control.
In the following sections we discuss an alternative approach 
to approximately solve (\ref{e-opt-ctrl}), 
with a long planning horizon $T$,
in a manner suitable for real-time, embedded control.
We will do this by moving most of the computational burden offline
(\ie, to controller synthesis, not execution),
using value function approximation.

\subsection{Value functions}
\label{s-value-functions}

We will use a standard trick in dynamic programming to express
the problem \eqref{e-opt-ctrl} as a similar one, with a shorter
time horizon $\tau<T$.
Define $V_\tau(z)$ as the optimal value of the problem
\begin{equation}
\label{e-value}
\begin{array}{ll}
\mbox{minimize} & 
  \sum_{t=\tau}^{T-1} g(\tilde x_t) + h(\tilde x_T) \\
\mbox{subject to} & \tilde x_{t+1} = f(\tilde x_t, \tilde u_t) \\
                  & \tilde x_\tau = z,
\end{array}
\end{equation}
with variables 
$u_\tau, \ldots, u_{T-1}$, $\tilde x_\tau, \ldots, \tilde x_T$.
Thus $V_\tau(z)$ is the optimal value of the control 
problem \eqref{e-opt-ctrl},
starting from state $z$ at time $\tau$.
It is also called the (Bellman) value function.


We can use the value function to rewrite \eqref{e-opt-ctrl} as
\begin{equation}
\label{e-one-step-policy}
\begin{array}{ll}
\mbox{minimize} & 
  \sum_{t=0}^{\tau - 1} g(\tilde x_t) + V_\tau(\tilde x_\tau) \\
\mbox{subject to} & \tilde x_{t+1} = f(\tilde x_t, \tilde u_t) \\
                  & \tilde x_0 = z,
\end{array}
\end{equation}
with variables $\tilde x_0, \dots, \tilde x_{\tau}$ and 
$\tilde u_0, \dots, \tilde u_{\tau-1}$.

Note that the new problem \eqref{e-one-step-policy}
has the same form as the MPC problem \eqref{e-opt-ctrl},
but with shorter horizon length $\tau$ and terminal cost $V_\tau$.
We can solve it by evaluating the cost of $K^\tau$ input sequences,
a far smaller number than the original problem, which requires evaluating
$K^T$ input sequences.
In the general case this advantage is an illusion, since 
the cost of evaluating $V_\tau(\tilde x_\tau)$ is $K^{T-\tau}$ 
evaluations of value input sequences.
But if the value function $V_\tau$ were known explicitly,
with a simple form, solving \eqref{e-one-step-policy}
would be easier than solving (\ref{e-opt-ctrl}).
For example, if $n$ is very small, say, 2 or 3, we could evaluate 
$V_\tau$ on a fine grid of points offline and store the values
in a table.

If we take $\tau=1$, we have
\begin{align*}
\phi_{\rm mpc}(z) = \argmin_{u} V_1 \big(f(z,u)\big).
\end{align*}
In this case, the control policy can be carried out 
using just $K$ evaluations of $V_1$ and $f$.
If $V_1$ were easy to evaluate (which it is not in general), this 
would give a very simple way to evaluate the MPC policy.


\subsection{Approximate value functions}
Unfortunately, computing, storing, and evaluating $V_\tau$
is difficult in all but a few special cases.
We therefore settle for an approximate or surrogate value 
function $\hat V_\tau$ which approximates $V_\tau$.
We consider the problem
\begin{equation}
\begin{array}{ll}
\mbox{minimize} & 
  \sum_{t=0}^{\tau-1} g(\tilde x_t) + \hat V_\tau (\tilde x_\tau) \\
\mbox{subject to} & \tilde x_{t+1} = f(\tilde x_t, \tilde u_t) \\
                  & \tilde x_0 = z,
\end{array}
\label{e-multi-step-policy}
\end{equation}
which is an approximation to \eqref{e-one-step-policy} 
with the same decision variables.
This optimization problem is the basis for the approximate MPC (A-MPC)
policy, $\phi_{\rm ampc}$.
We define $\phi_{\rm ampc}(z)$ as an optimal value of $\tilde u_0$
for \eqref{e-multi-step-policy}.
If we have $\hat V_\tau = V_\tau$
(\ie, $\hat V_\tau$ is a perfect approximation),
we recover the MPC policy,
regardless of the shortened horizon length $\tau$.
Alternatively, if we take $\tau=T$ and $\hat V_T(z) = h(z)$, 
we recover the MPC policy.

\section{Obtaining an approximate value function}
\label{s-deriving-value-function}
In order to use an approximate MPC policy,
we first need to synthesize an approximate value function $\hat V_\tau$.
In practice, good control can be obtained with very simple forms for $\hat V_\tau$,
even when $\hat V_\tau$ is not a particularly good estimate of $V_\tau$.
(Indeed, the success of short-horizon predictive control 
suggests that even a greedy policy, obtained using $\hat V_\tau = g$, works well
in some cases.)

\paragraph{Quadratic functions.}
We limit our attention to quadratic value function approximations,
\ie, functions of the form
\begin{align}
\label{e-quad-value-fun}
\hat V_\tau(x) = (x - x_{\rm des})^T P (x - x_{\rm des}) + r,
\end{align}
where $P \in \reals^{n\times n}$ is a symmetric positive definite matrix,
$r\in \reals$ is a scalar,
and
$x_{\rm des}$ is a desired system state,
representing an area of the state space with low cost.
The scalar $r$ has no effect on the A-MPC control policy, but is included
since we will choose it and $P$ so that $\hat V_\tau \approx V_\tau$.

For example, one simple 
quadratic approximate value function is given by
\begin{align}
\label{e-vhat-energy}
\hat V_\tau(x) = (x - x_{\rm des})^T P_{\rm energy} (x - x_{\rm des}),
\end{align}
where the symmetric matrix $P_{\rm energy}$ is such that
$x^T P_{\rm energy} x$ gives the total energy stored in the converter.
Control policies based on the energy function appear
to have good stability properties;
for details, see \cite{sanders1992lyapunov} and \cite{midya1992optimal}.



\paragraph{Function fitting.}
\label{s-function-fitting}
We first choose state vectors $x^{(1)}, \dots, x^{(N)}$,
which are points at which we desire an accurate approximation.
Then for each state vector $x^{(i)}$, we compute the value function
$V_\tau(x^{(i)})$ by solving problem \eqref{e-opt-ctrl} with $z = x^{(i)}$.
This is done offline, possibly using substantial computational resources.
We then choose the parameters $P$ and $r$ of the approximate value function
as the solution to the least-squares approximation problem
\begin{equation}
\begin{array}{ll}
\mbox{minimize} & 
  \displaystyle \frac1N
  \sum_{i=1}^N \Big( V_\tau\big(x^{(i)}\big) - \hat V_\tau\big(x^{(i)}\big) \Big)^2
  +
  \lambda(P - \alpha P_{\rm energy})^2.
\end{array}
\label{e-approx-problem}
\end{equation}
The decision variables are the approximate value function parameters
$P$ and $r$, as well as the scaling factor $\alpha$.
Note that $P$ and $r$ enter linearly in $\hat V_\tau$,
which allows us to solve this problem as a least-squares problem.
The parameter $\lambda>0$ is a regularization parameter;
if we choose $\lambda$ to be small,
the optimal $P$ and $r$ almost minimize
mean-squared error between the approximate value function
and the sampled values,
thus returning a good value function approximation at the points.
By choosing $\lambda$ large,
we force $P$ to be close to (a scaled version of) $P_{\rm energy}$;
this tends to make the resulting control more robust,
especially if we have few samples.

With values of $P$ and $r$ that solve \eqref{e-approx-problem},
we expect that $\hat V_\tau(z) \approx V_\tau(z)$ 
in areas of the state space that are well represented by the state samples.
Because the flexibility (and therefore, accuracy)
of the approximate value function is a limited resource,
it is important to chose state samples that are likely
to be encountered in normal operation,
and not to include samples
far from relevant areas of the state space,
since including them may decrease accuracy in relevant areas.

We can explicitly add the constraints that $P$ be positive definite,
and $\alpha \geq 0$; the approximation problem above is no longer
a least-squares problem but it is still convex and readily solved
\cite{boyd2004convex}.
But when the parameter $\lambda$ and the sample points are appropriately
chosen, these constraints are typically satisfied automatically.
We also note that since we are approximating the value function, we 
do not need to evaluate $V_\tau(x^{(i)})$ exactly; good approximations are
good enough.

\section{Evaluating the approximate policy}
\label{s-implementation}
In this section we first show how to evaluate the A-MPC policy in general;
we then discuss several special cases
in which this process can be simplified,
often with great computational benefits.
We also discuss how to (approximately) handle switching costs.

\subsection{General $f$ and $g$}
To carry out the approximate MPC policy,
we first measure the current system state.
We then solve \eqref{e-multi-step-policy},
with $z$ set to the current state,
by first finding the state trajectory 
(simulating the system for each of the $K^\tau$ different input trajectories),
and evaluating the cost function for each simulated trajectory.
Once we have obtained the input trajectory that attains the lowest cost,
we apply the optimal first control input $\tilde u_0$.
The process is then repeated at the next sample time.
This is the same process that can be used to solve the MPC problem \eqref{e-opt-ctrl}
when the horizon length is short enough,
and reflects the fact that 
\eqref{e-opt-ctrl} and \eqref{e-multi-step-policy}
have the same form.

Evaluating the value function requires finding a full desired state $x_{\rm des}$.
This contrasts with the MPC problem (\ref{e-opt-ctrl}),
in which a full desired state is not required,
and often only a few (but not all) state variables appear in the cost function.
One might ask why state variables appear in the value function
that do not appear in any cost function.
This is because, although some state variables do not appear explicitly
in the value function, 
they can play an important role in the future evolution of the system,
so that deviation of these state variables from `good' values
may result in high cost in the future,
\ie, result in deviation of other, explicitly penalized variables from `good' values.

For example, for the boost converter example in \S\ref{s-examples},
we explicitly penalize only output voltage deviation from a desired value,
and not inductor current deviation.
However, we expect the approximate value function to penalize
inductor current deviation,
because the inductor current has a substantial effect
on \emph{future} values of the output voltage;
a deviation of the inductor current from `good' values
results in a future deviation of capacitor voltage from the
desired target in the future.

Finding appropriate values of $x_{\rm des}$ is case specific,
but usually involves only simple computations;
in the boost converter example,
we take the desired inductor current to simply be equal to the load current
when the output voltage is equal to the desired value.

\paragraph{One-step horizon.}
In the case $\tau = 1$, we can simplify 
\eqref{e-multi-step-policy}
by eliminating the states as variables,
allowing us to write the policy as
\begin{align*}
\label{e-one-step-policy}
\phi_{\rm ampc}(x) = \argmin_{u} \hat V_1 \big(f(x,u)\big).
\end{align*}
In this case the approximate MPC policy requires only 
$K$ evaluations of $f$ and $V_1$.

\subsection{Switched-affine systems}
In this section we present a solution approach
for switched-affine dynamics and quadratic costs.
More specifically, we assume the system dynamics \eqref{e-sw-sys}
can be written as
\begin{align*}
x_{t+1} = A^{u_t}x_t + b^{u_t},
\end{align*}
for matrices $A^1, \ldots, A^K$ and vectors $b^1,\ldots, b^K$,
and that the stage costs $g$ are quadratic, so that
\[
g(x) = (x-x_{\rm des})^T Q (x-x_{\rm des}).
\]
where we have $x_{\rm des} = Cx + d$,
\ie, the desired state is an affine function of the current state.

In this case, 
the optimization problem \eqref{e-multi-step-policy}
can be written as:
\begin{equation}
\begin{array}{ll}
\mbox{minimize} &  
  z^T \tilde P(u) z + \tilde q(u)^T z + \tilde r(u). \\
\end{array}
\label{e-opt-ctrl-sw-aff}
\end{equation}
The variable is the vector 
$u = (\tilde u_{\tau-1}, \ldots, \tilde u_0)$.
The functions $\tilde P$, $\tilde q$, $\tilde r$
are given by

\begin{align*}
\tilde P(u) 
&= 
\sum_{t=0}^\tau 
x_0^T \Phi_{0,t}^T (I-C)^T P_t (I-C) \Phi_{0,t} x_0
\\
\tilde q(u) 
&= 
2 \sum_{t=0}^\tau e_t^T P_t (I-C) \Phi_{0,t} x_0
\\
\tilde r(u) 
&= 
\sum_{t=0}^\tau e_t^T P_t e_t
\\
e_t 
&= (I-C)\sum_{s=0}^{t-1} \Phi_{s+1,t} b^{u_s} - x_{\rm des}.
\\
\Phi_{s,t} 
&= 
\begin{cases}
0 & t < s \\
I & t = s \\
A^{t-1} \cdots A^{s} & t > s \\
\end{cases}.
\end{align*}
\begin{derivations}
We have
\begin{align*}
x_t 
&= 
\left( A^{t-1} \cdots A^0 \right)x_0
+ \sum_{\tau=0}^{t-1} \left( A^{t-1} \cdots A^{\tau+1} \right) b^\tau
\\&= \Phi_{0,t} x_0 + \sum_{\tau=0}^{t-1} \Phi_{\tau+1,t} b^\tau
\end{align*}
with
\[
\Phi_{\tau,t} = 
\begin{cases}
0 & t < \tau \\
I & t = \tau \\
A^{u_{t-1}} \cdots A^{u_{\tau}} & t > \tau \\
\end{cases}.
\]

The total cost is
\begin{align*}
&\sum_{t=0}^T (x_t - x_{\rm des})^T Q (x_t - x_{\rm des})
\\&=
\sum_{t=0}^T ((I - C)x_t - d)^T Q ((I - C)x_t - d)
\\&=
\sum_{t=0}^T 
\left( (I-C) \left( \Phi_{0,t} x_0 + \sum_{\tau=0}^{t-1} \Phi_{\tau+1,t} b^{u_\tau} \right) - d \right)^T
Q 
\left( (I-C) \left( \Phi_{0,t} x_0 + \sum_{\tau=0}^{t-1} \Phi_{\tau+1,t} b^{u_\tau} \right) - d \right)
\\&=
\sum_{t=0}^T 
\left( (I-C) \Phi_{0,t} x_0 + (I-C)\sum_{\tau=0}^{t-1} \Phi_{\tau+1,t} b^{u_\tau}  - d \right)^T
Q 
\left( (I-C) \Phi_{0,t} x_0 + (I-C)\sum_{\tau=0}^{t-1} \Phi_{\tau+1,t} b^{u_\tau}  - d \right)
\\&=
\sum_{t=0}^T 
\left( (I-C) \Phi_{0,t} x_0 + e_t \right)^T
Q 
\left( (I-C) \Phi_{0,t} x_0 + e_t \right)
\\&=
\sum_{t=0}^T 
x_0^T
\Phi_{0,t}^T
(I-C)^T
Q 
(I-C) \Phi_{0,t} x_0
+
2 \sum_{t=0}^T e_t^T Q (I-C) \Phi_{0,t} x_0
+ \sum_{t=0}^T e_t^T Q e_t
\end{align*}
where
\[
e_t = 
(I-C)\sum_{\tau=0}^{t-1} \Phi_{\tau+1,t} b^{u_\tau} - x_{\rm des}.
\]
\end{derivations}
We use the shorthand $P_t = Q$ if $t < \tau$,
and $P_t = P$ if $t = \tau$.

We can precompute the parameters of these functions
for each of the $K^\tau$ different values of $u$.
To solve \eqref{e-opt-ctrl-sw-aff},
we evaluate each of the $K^\tau$ different quadratic functions at $z$
and note which quadratic function attains the lowest value;
the corresponding index $u$ is a solution to 
\eqref{e-opt-ctrl-sw-aff}.
By precomputing $\tilde P$, $\tilde q$, and $\tilde r$,
we can acheive significant computational savings
compared to a simulation-based approach.

\subsection{Linear systems with switched input}
We consider a special case of the switched-affine systems.
We assume quadratic stage costs and dynamics of the form
\begin{align}
\label{e-lin-sys-sw-input}
x_{t+1} = Ax_t + b^{u_t}.
\end{align}
Because this is a special case of the switched affine case above,
the A-MPC problem \eqref{e-opt-ctrl}
can also be written as \eqref{e-opt-ctrl-sw-aff}.
However, because the parameter $\tilde P$ does not depend on the control input $u$,
it can be dropped from the optimization problem 
without affecting the optimal decision variables,
so that the A-MPC problem \eqref{e-opt-ctrl} is equivalent to
\begin{equation}
\begin{array}{ll}
\mbox{minimize} &  
  \tilde q(u)^T z + \tilde r(u). \\
\end{array}
\label{e-opt-ctrl-sw-lin}
\end{equation}
The variable is the vector 
$u = (\tilde u_{\tau-1}, \ldots, \tilde u_0)$.
The functions $\tilde q$, $\tilde r$ are given above.
As before, we can precompute the parameters of the affine functions
in the objective of \eqref{e-opt-ctrl-sw-lin}
for each of the $K^\tau$ different values of $u$.


\paragraph{One-step horizon.} 
If we have $\tau = 1$ in addition to 
linear systems with switched inputs and quadratic costs,
then (\ref{e-multi-step-policy}) can be written as
\begin{align*}
\phi_{\rm ampc}(x) = \argmin Fx + g.
\end{align*}
In this case we interpret $\argmin$ 
to give minimum over all indices of the vector $Fx + g$.
The matrix $F \in \reals^{K \times n}$ and 
the vector $g \in \reals^K$ can be precomputed as
\[
F =
\begin{bmatrix}
2b^{1T} P A^1 \\
\vdots \\
2b^{KT} P A^K \\
\end{bmatrix},
\qquad
g =
\begin{bmatrix}
(b^1 - x_{\rm des})^T P (b^1 - x_{\rm des}) \\
\vdots \\
(b^K - x_{\rm des})^T P (b^K - x_{\rm des})
\end{bmatrix}.
\]
Note that if $x_{\rm des}$ changes during the operation of the converter,
only $g$ needs to be updated.

\section{Summary of approximate MPC method}
Here we summarize the steps needed to use the approximate MPC policy.

First, the parameters of the underlying MPC policy are determined.
This means that a dynamics model \eqref{e-sw-sys}
of the system is derived and
the cost function and horizon length are determined.
The dynamics model should be simple enough, and the horizon length modest enough,
for problem \eqref{e-opt-ctrl}
to be solved fairly quickly on a desktop computer,
so that many (hundreds or thousands) of samples can be generated reasonably quickly.

Second,
the reduced horizon length $\tau$ is chosen,
based on the computational capabilities of the embedded control platform
(\ie, a large value of $\tau$ 
results in a more accurate approximation of the MPC problem,
but also in a higher computational burden,
as the controller will need to evaluate $K^\tau$ different trajectories each time step.)
Sample states likely to be encountered in normal operation are then chosen,
and the value function $V_\tau(x^{(i)}) = V_\tau(x^{(i)})$
is evaluated for each sample $x^{(i)}$ by solving \eqref{e-value}.
Once the value function has been evaluated for all sample points,
we choose the regularization tradeoff parameter $\lambda$
and solve the approximation problem \eqref{e-approx-problem}
to obtain the approximate value function
parameters $P$ and $r$.

Once the approximate value function is obtained,
the control policy can be evaluated by solving the short-horizon policy 
\eqref{e-multi-step-policy},
implementing the optimal first control input $u_0$,
waiting until the next sampling time, and repeating,
as discussed in \S\ref{s-implementation}.

\section{Extensions}
\subsection{Switching cost}
In this section, we describe how to include switching costs,
\ie, a cost associated with switching from one control input to another,
so that (\ref{e-opt-ctrl}) becomes
\begin{equation}
\begin{array}{ll}
\mbox{minimize} & 
  \sum_{t=0}^{T - 1} \big( g(\tilde x_t) + \ell(u_{t-1}, u_t) \big)
  + h(\tilde x_{T}) \\
\mbox{subject to} & \tilde x_{t+1} = f(\tilde x_t, \tilde u_t) \\
                  & \tilde x_0 = z.
\end{array}
\label{e-opt-ctrl-switched}
\end{equation}
where $\ell$ is a switching cost function,
and the parameter $u_{t-1}$ represents the control input
most recently implemented.
In theory,
this problem
can be put into the form of problem (\ref{e-opt-ctrl})
by encoding the switch position as a state variable.

However, a more convenient way to handle the switching cost is to 
include it in the first $\tau$ time steps 
and simply ignore it during the last $T-\tau$ time steps.
By doing this, the process of deriving the value function 
(see \S\ref{s-deriving-value-function})
remains the same,
but the control policy is modified to reflect the switching costs:
\begin{equation}
\begin{array}{ll}
\mbox{minimize} & 
  \sum_{t=0}^{\tau-1} 
  \big( g(\tilde x_t) + \ell(u_{t-1}, u_t) \big) + \hat V_\tau (\tilde x_\tau) \\
\mbox{subject to} & \tilde x_{t+1} = f(\tilde x_t, \tilde u_t) \\
                  & \tilde x_0 = z.
\end{array}
\label{e-multi-step-policy-switching}
\end{equation}
Trivial modifications to this setup allow us to include
many of the standard tricks for FCS-MPC can be applied directly,
such as using a frequency-based switching cost
to enforce a constant switching frequency.
(See \cite{kouro2009model, cortes2008predictive}, for details.)
We expect this approach to work well for
converters in which the switching dynamics do not
have a significant long-term effect,
\eg, converters with a high switching frequency.

\section{Examples}
\label{s-examples}

\subsection{Boost converter example}


\label{s-boost-example}
\begin{figure}
\centering
\resizebox{.45\textwidth}{!}{%

\begin{tikzpicture}[scale=1.1]
\tikzstyle{every node}=[font=\huge]
\tikzstyle motor=[color=gray!50!white,draw=black,thick]
\ctikzset{bipoles/diode/height=.375}
\ctikzset{bipoles/diode/width=.3}

\draw[scale=1] (0,0) 
to [L, l=$L_{}$] ++(2,0)
to [R, l=$R_L$] ++(2,0) node(N_Q){}
to [full diode] ++(3,0) node(N_C){}
-- ++(2,0) node(N_load){}
to [R, l=$R_{\rm load}$] ++(0,-3)
-- (0, -3)
to [american voltage source, l=$V_{\rm dc}$] ++(0,3)
;

\draw[scale=1] 
(N_Q) ++ (0,-3) to [cspst, bipoles/length=3cm, l=$u$] (N_Q)
;

\draw[scale=1] 
(N_C) to [C, l_=$C$] ++(0,-3)
;

\end{tikzpicture}
}
\caption{
Boost converter model.
}
\label{f-boost}
\end{figure}
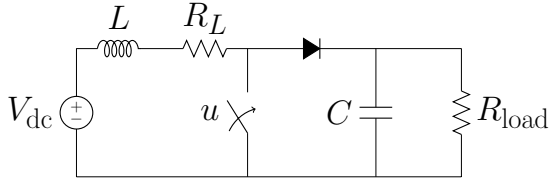

We consider the ideal boost converter shown in figure~\ref{f-boost}.
The system state is
\[
x_t =
\begin{bmatrix} i_{L,t} \\ v_{C,t} \end{bmatrix}.
\]
where $i_{L,t}$ and $v_{C,t}$ are the inductor current and capacitor
voltage at discrete epoch $t$.
We (approximately) discretize the converter dynamics
to have the form \eqref{e-sw-sys}.
The details of the discretization are given appendix \ref{s-boost-derivation}.

We use the cost function to 
penalize deviation of the capacitor voltage from a desired value:
\[
g(x_t) = |v_{C,t} - v_{\rm des}|.
\]
With this cost function, 
and with the discretized dynamics, 
problem \eqref{e-opt-ctrl}  can be written as a mixed-integer linear program (MILP).

We used the parameters of \cite{karamanakos2013direct},
so that
$V_{\rm dc} = 10$ V,
$L = 450$ $\mu$H,
$R_L = 0.3$ $\Omega$,
$C = 220$ $\mu$F,
and
$R_{\rm load} = 73$ $\Omega$.
We took $v_{\rm des} = 30$ V
(instead of $v_{\rm des} = 15$ V, as in \cite{karamanakos2013direct}),
because this increased target voltage means that the controller must
charge up the inductor at the cost of temporarily increasing voltage error.
(For $v_{\rm des} = 15$ V,
the controller can simply turn off the switch,
charging the inductor and capacitor simultaneously
until the output voltage reaches the desired value.)

\begin{figure}
\centering
\psfrag{V}[bc]{\small $V$}
\psfrag{v}[cr]{\small $v$ ($\rm V$)}
\psfrag{i}[bl]{\small $i$ ($\rm A$)}
\ifarxiv
\includegraphics[width=.4\textwidth]{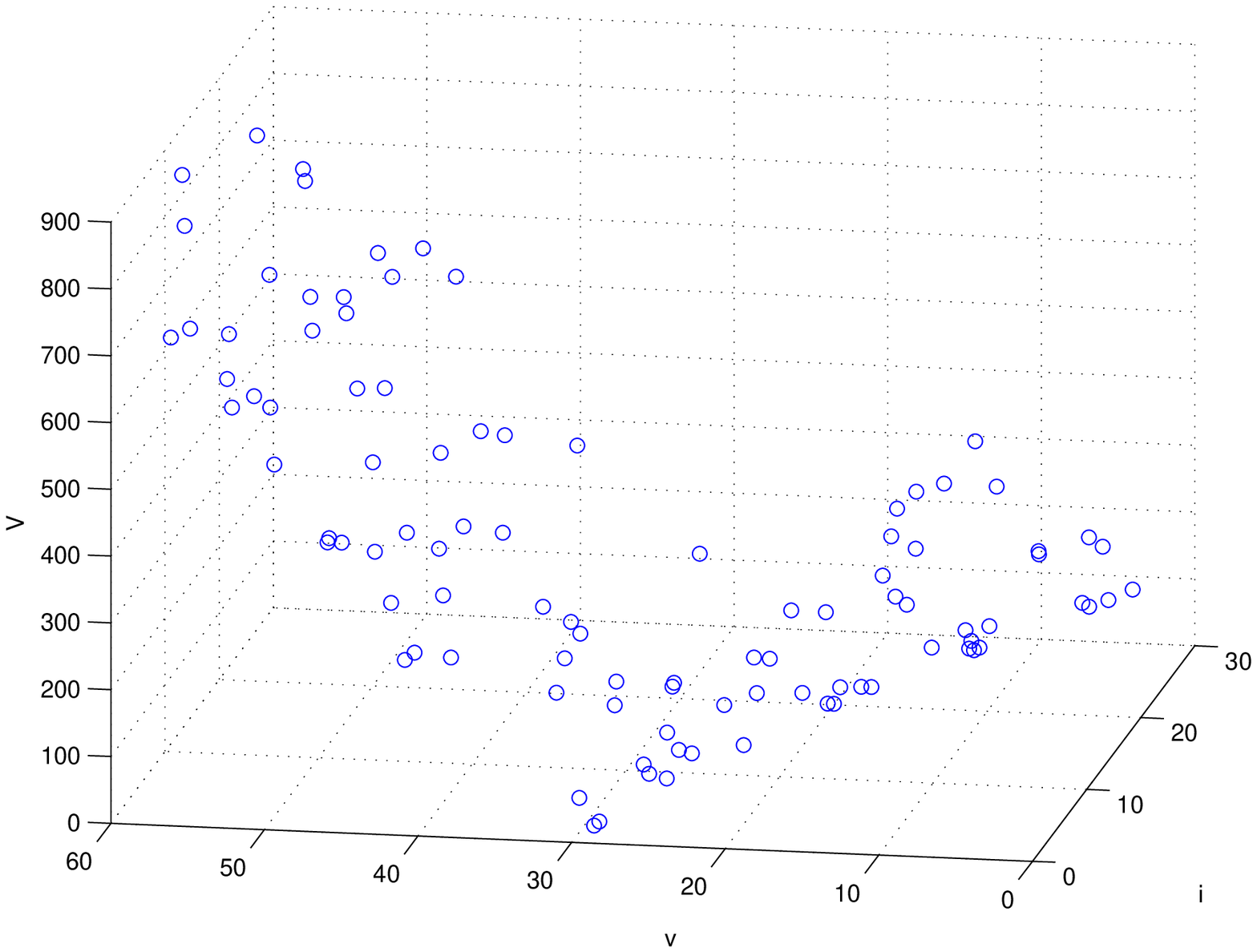}
\else
\includegraphics[width=.4\textwidth]{jl/boost_V.eps}
\fi
\qquad\qquad
\psfrag{V}[bc]{\small $\hat V_\tau$}
\ifarxiv
\includegraphics[width=.4\textwidth]{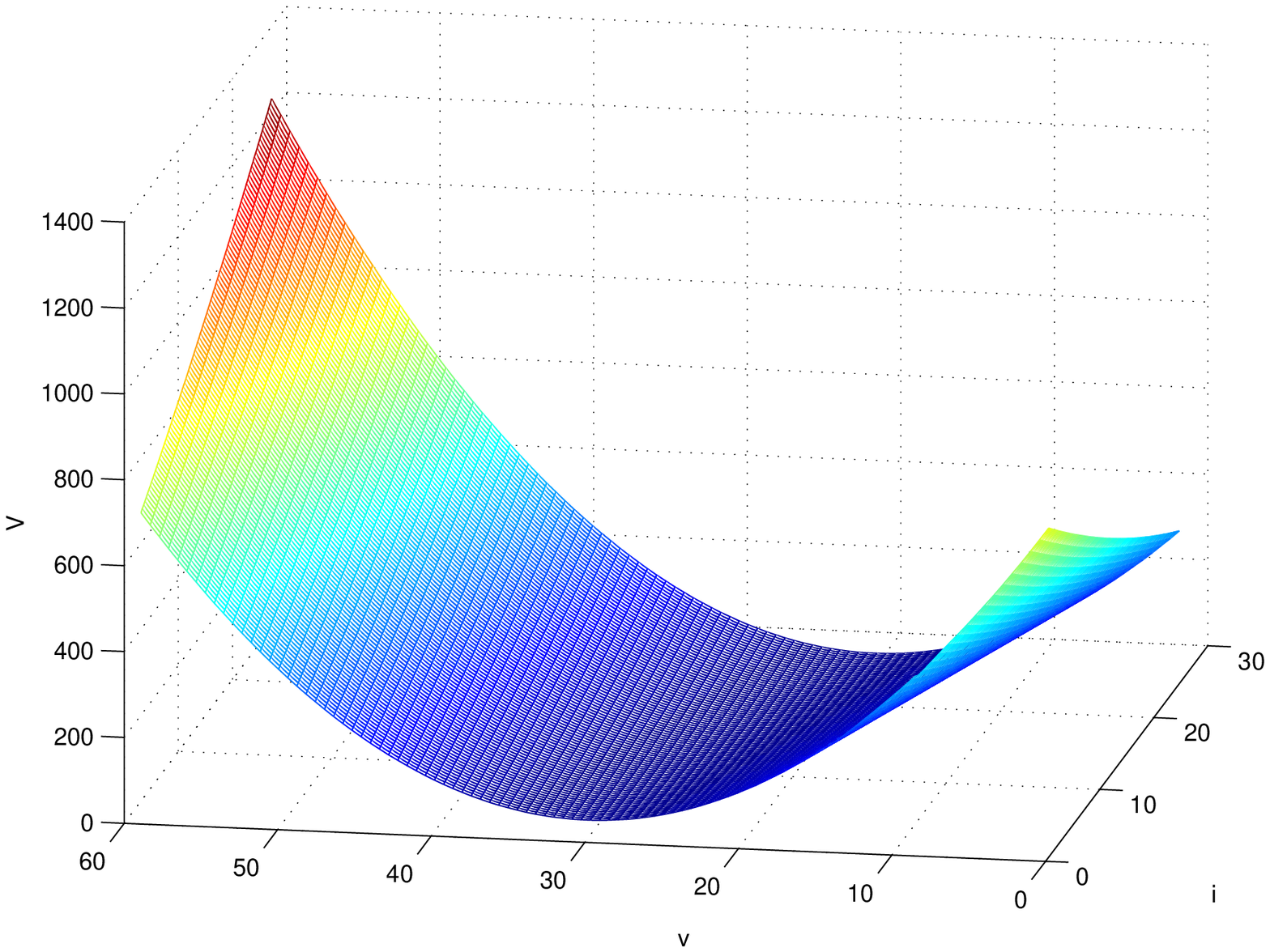}
\else
\includegraphics[width=.4\textwidth]{jl/boost_Vhat.eps}
\fi
\caption{
For the boost converter example,
the $100$ sampled points of the value function are shown on the left,
and the fitted quadratic approximate value function $\hat V_\tau$
is shown on the right,
as a function of the inductor current $i$
and the capacitor voltage $v$.
}
\label{f-V}
\end{figure}

Following \S\ref{s-function-fitting},
we randomly generated $100$ initial states
by sampling uniformly over a box with
a lower bound of $(0\; \rm{A}, 0\; \rm{V})$
and an upper bound of $(10\; \rm{A}, 50\; \rm{V})$,
which are reasonable limits for normal operation of the converter.
We then solved problem \eqref{e-opt-ctrl} for each of the initial states,
to within one percent accuracy.
Using these sample points,
we solved the regularized approximation problem 
\eqref{e-approx-problem}
to obtain an approximate value function,
with parameter $\lambda = 100$, 
using a positive semidefinite constraint on $P$.
The resulting value function is shown in figure \ref{f-V}
along with the $100$ randomly sampled values.
To evaluate the ADP policy, 
we must explicitly define a desired state vector $x_{\rm des}$,
which we take to be $(v_{\rm des} / R_{\rm load}, v_{\rm des}$).

\begin{figure}
\centering
\psfrag{v}[bc]{Cap.\ voltage ($\rm V$)}
\psfrag{i}[bc]{Ind.\ current ($\rm A$)}
\psfrag{t}[c]{time ($\rm s$)}
\ifarxiv
\includegraphics[width=.75\textwidth]{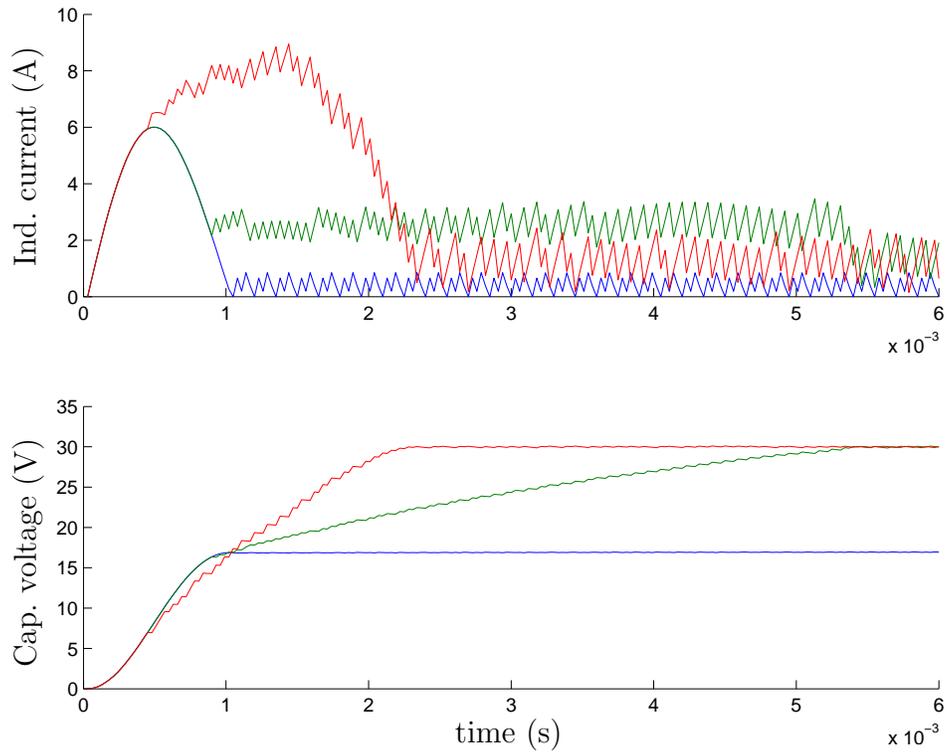}
\else
\includegraphics[width=.75\textwidth]{jl/boost_mpc.eps}
\fi
\caption{
Closed-loop trajectories for the boost converter example with the MPC policy,
for different values of the planning horizon $T$.
The desired output voltage value is $30$ V.
}
\label{f-traj-mpc}
\end{figure}

In figure \ref{f-traj-mpc} we show the closed-loop response of the system
under FCS-MPC controllers with horizon lengths of $1$, $5$, and $30$,
with initial condition $x_0 = (0, 0)$ 
(\ie, zero inductor current and capacitor voltage).
To evaluate the FCS-MPC control policy,
(\ref{e-opt-ctrl}) was solved to one percent accuracy.
With a horizon length of $T = 1$,
the controller does not reach the desired steady-state
value of $30$ V
and for $T=5$, the desired steady-state value is reached,
but only slowly.
For $T =30$, the desired steady-state value is reached very quickly;
however, using such a long planning horizon is not practical.
(In our desktop implementation, 
each MPC evaluation took around 10 minutes on average.)

\begin{figure}
\centering
\psfrag{v}[bc]{Cap.\ voltage ($\rm V$)}
\psfrag{i}[bc]{Ind.\ current ($\rm A$)}
\psfrag{t}[c]{time ($\rm s$)}
\ifarxiv
\includegraphics[width=.75\textwidth]{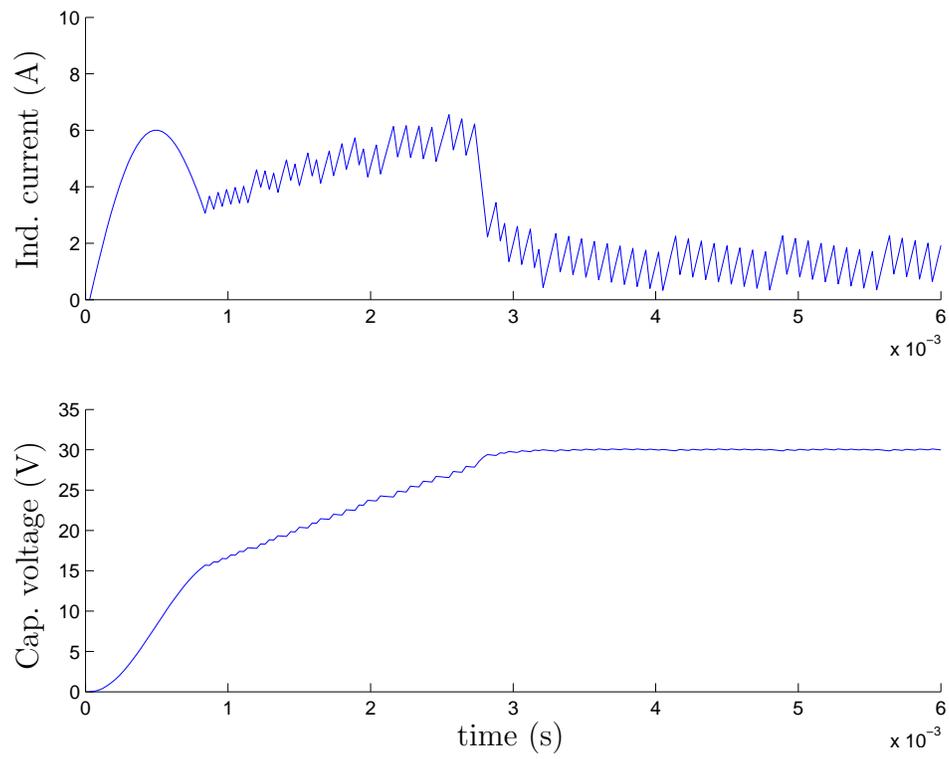}
\else
\includegraphics[width=.75\textwidth]{jl/boost_adp.eps}
\fi
\caption{
Closed-loop trajectories for the boost converter example,
using the A-MPC policy.
The desired output voltage value is $30$ V.
}
\label{f-traj}
\end{figure}

In figure \ref{f-traj},
we show the response using
the A-MPC controller, using $\tau = 1$ and using the
approximate value function derived from function fitting.
We can see that the A-MPC controller outperforms the
MPC controllers with short horizons ($T = 1$ and $T =5$),
and is on par with the long-horizon MPC policy ($T=30$).

\subsection{Inverter example}
\label{s-inverter-example}
In this example, we consider a three-phase inverter
with an LCL filter.
The input is a constant DC source,
and the output is a three-phase AC voltage load.
The goal is to maintain a sinusoidal output current waveform
with a desired amplitude and a high power factor.

\begin{figure}
\centering
\resizebox{.9\textwidth}{!}{%

\def\bridgewidth{1.5}
\def\bridgeswheight{2}
\def\bridgevertsep{.5}
\def\bridgetoLCL{2}
\def\LCLheight{2.5}
\def\capsep{2}
\def\capdistance{.5}
\def\sourcebridgesep{2}

\begin{tikzpicture}[scale=1.1]
\tikzstyle{every node}=[font=\LARGE]
\tikzstyle motor=[color=gray!50!white,draw=black,thick]
\ctikzset{bipoles/diode/height=.375}
\ctikzset{bipoles/diode/width=.3}

\draw[scale=1] (0,0) 
-- ++(\sourcebridgesep, 0) node(N_bridge1top){}
-- ++(\bridgewidth, 0) node(N_bridge2top){}
-- ++(\bridgewidth, 0) node(N_bridge3top){}
;

\draw[scale=1] (N_bridge1top)
to [cspst, bipoles/length=3cm] ++(0,-\bridgeswheight) node(N_bridge1center){}
-- ++(0,-1)
to [cspst, bipoles/length=3cm] ++(0,-\bridgeswheight)
;

\draw[scale=1] (N_bridge2top)
to [cspst, bipoles/length=3cm] ++(0,-\bridgeswheight)
-- ++(0,-.5) node(N_bridge2center){}
-- ++(0,-.5)
to [cspst, bipoles/length=3cm] ++(0,-\bridgeswheight)
;

\draw[scale=1] (N_bridge3top)
to [cspst, bipoles/length=3cm] ++(0,-\bridgeswheight)
-- ++(0,-1) node(N_bridge3center){}
to [cspst, bipoles/length=3cm] ++(0,-\bridgeswheight) node(N_bridge3bottom){}
;

\draw[scale=1] (N_bridge3bottom)
-- ++(-2*\bridgewidth-\sourcebridgesep,0)
to [american voltage source, l=$V_{\rm dc}$] ++(0,2*\bridgeswheight+2*\bridgevertsep)
;

\draw[scale=1] (N_bridge1center) -- ++(0,0)
-- ++(0,0)
-- ++(2*\bridgewidth+\bridgetoLCL,0)
-- ++(0,\LCLheight/2 -\bridgevertsep)
to [L, l=$L_{1}$, ] ++(3,0) node(N_cap1top){}
-- ++(2*\capsep, 0)
to [L, l=$L_{2}$] ++(3,0)
to [sinusoidal voltage source] ++(3,0)
-- ++(0,-\LCLheight)
;

\draw[scale=1] (N_bridge2center) -- ++(0,0)
-- ++(\bridgewidth+\bridgetoLCL,0)
to [L, l=$L_{1}$] ++(3,0)
-- ++(\capsep, 0) node(N_cap2top){}
-- ++(\capsep, 0)
to [L, l=$L_{2}$] ++(3,0)
to [sinusoidal voltage source] ++(3,0)
;

\draw[scale=1] (N_bridge3center) -- ++(0,0)
-- ++(\bridgetoLCL,0)
-- ++(0,-\LCLheight/2 +\bridgevertsep)
to [L, l=$L_{1}$] ++(3,0)
-- ++(2*\capsep, 0) node(N_cap3top){}
to [L, l=$L_{2}$] ++(3,0)
to [sinusoidal voltage source] ++(3,0)
;

\draw[scale=1] (N_cap1top) -- ++(0,0)
-- ++(0, -\LCLheight)
-- ++(0, -\capdistance)
to [C, l_=$C$] ++(0,-1.5)
-- ++(2*\capsep,0)
;

\draw[scale=1] (N_cap2top) -- ++(0,0)
-- ++(0, -\LCLheight/2)
-- ++(0, -\capdistance)
to [C, l_=$C$] ++(0,-1.5)
;

\draw[scale=1] (N_cap3top) -- ++(0,0)
-- ++(0, -\capdistance)
to [C, l_=$C$] ++(0,-1.5)
;

\end{tikzpicture}
}
\caption{
Inverter model.
}
\label{f-inverter}
\end{figure}
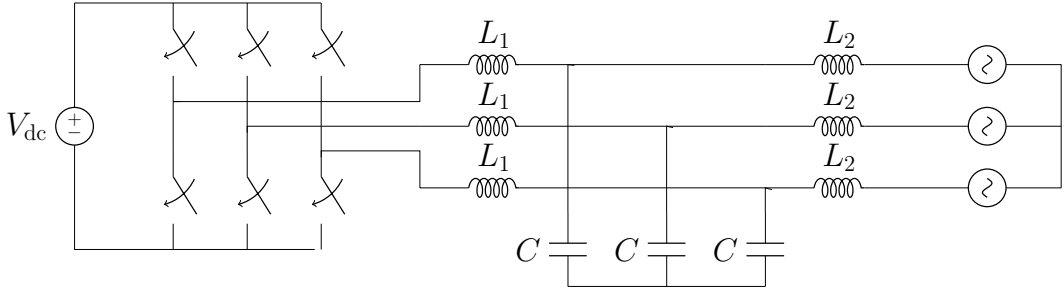

\paragraph{Model.}
The inverter is shown in figure~\ref{f-inverter}.
The AC output voltage signals are sinusoidal,
each with amplitude of $V_{\rm load}$
and with a phase difference of $\frac{2\pi}{3}$ between them.
The control input to the system is the positions of the six switches;
we assume the two switches in each bridge cannot both be open or closed.

The inverter dynamics can be expressed in the form of (\ref{e-lin-sys-sw-input}),
\ie, as a linear system with a switched input.
For the full derivation of the dynamics,
see appendix \ref{s-inverter-derivation}.
We note here that to put the dynamics in this form,
we include in the state vector not only the
currents of the six inductors
and the voltages of the three capacitors,
but also the values $\sin \omega t$ and $\cos \omega t$.
These extra state variables are used
to express both the time-varying output voltage signal and desired output current signals
in a linear, time-invariant form.

We use the cost function to 
penalize deviation of the load currents from desired values:
\[
g(x_t) = 
\left |i_{4,t} - I_{\rm des} \sin \omega t \right | +
\left |i_{5,t} - I_{\rm des} \sin \left(\omega t - \textstyle\frac{2\pi}{3}\right) \right | +
\left |i_{6,t} - I_{\rm des} \sin \left(\omega t - \textstyle\frac{4\pi}{3}\right) \right |.
\]
where $I_{\rm des}$ is the desired peak current value,
and 
$i_{4,t}$,
$i_{5,t}$,
and
$i_{6,t}$
are the values of the output currents
(\ie, the currents through the inductors on the righthand side of figure~\ref{f-inverter}).
Because each of the three sinusoidal terms can be written
as a linear combination of $\sin \omega t$ and $\cos \omega t$,
including these values as state variables
allows us to write the righthand side in terms of the state vector,
independent of $t$.
With this cost function, 
and with the discretized dynamics given in appendix \ref{s-inverter-derivation},
problem \eqref{e-opt-ctrl}  can be written as a mixed-integer quadratic program (MIQP).

We used the parameters 
$V_{\rm dc} = 700$ V,
$L_1 = 6.5$ $\mu$H,
$L_2 = 1.5$ $\mu$H,
$C = 15$ $\mu$F,
$V_{\rm load} = 300$ V,
and
$I_{\rm des} = 10$ A.


\paragraph{ADP controller.}
Following \S\ref{s-function-fitting},
we randomly generated $1000$ initial states
by sampling uniformly over a box.
The bounds of the box were chosen to be 
$-20$ A and $20$ A for state variables representing inductor currents,
$-300$ V and $300$ V for state variables representing capacitor voltages,
and $-1$ and $1$ for the state variables $\sin \omega t$ and $\cos \omega t$.
We solved problem \eqref{e-opt-ctrl} for each of the initial states,
to within one-percent accuracy.
Using these sample points,
we then solved the regularized approximation problem 
\eqref{e-approx-problem}
to obtain an approximate value function,
with parameter $\lambda = 1$ 
and no constraints on $P$ or $\alpha$.
The energy corresponding of the terms 
$\sin \omega t$ and $\cos \omega t$,
used to define the matrix $E$ in (\ref{e-approx-problem}),
was taken to be $0$.

When evaluating the controllers 
(using (\ref{e-multi-step-policy-switching})),
we incorporate a switching cost into the control.
We use a constant switching cost,
so that $\ell(u_{t-1}, u_t) = 0$ if $u_{t-1} = u_t$,
and $\ell(u_{t-1}, u_t) = 1$ otherwise.

To use (\ref{e-quad-value-fun}),
we must find a desired state vector $x_{\rm des}$,
as an affine function of the current state.
Note that for a given value of $\sin\omega t$ and $\cos \omega t$,
there is a unique set of load currents 
$i_{4,t}$,
$i_{5,t}$,
and
$i_{6,t}$
that result in zero cost.
Corresponding other inductor currents and capacitor voltages
are found using phasor calculations,
and expressed as an affine function of the state variables
(namely $\sin \omega t$ and $\cos \omega t$).

\begin{figure}
\centering
\psfrag{ii}[bc]{Input curr.\ ($\rm A$)}
\psfrag{v}[bc]{Cap.\ voltage ($\rm V$)}
\psfrag{io}[bc]{Output curr.\ ($\rm A$)}
\psfrag{t}[cc]{time ($\rm s$)}
\ifarxiv
\includegraphics[width=.85\textwidth]{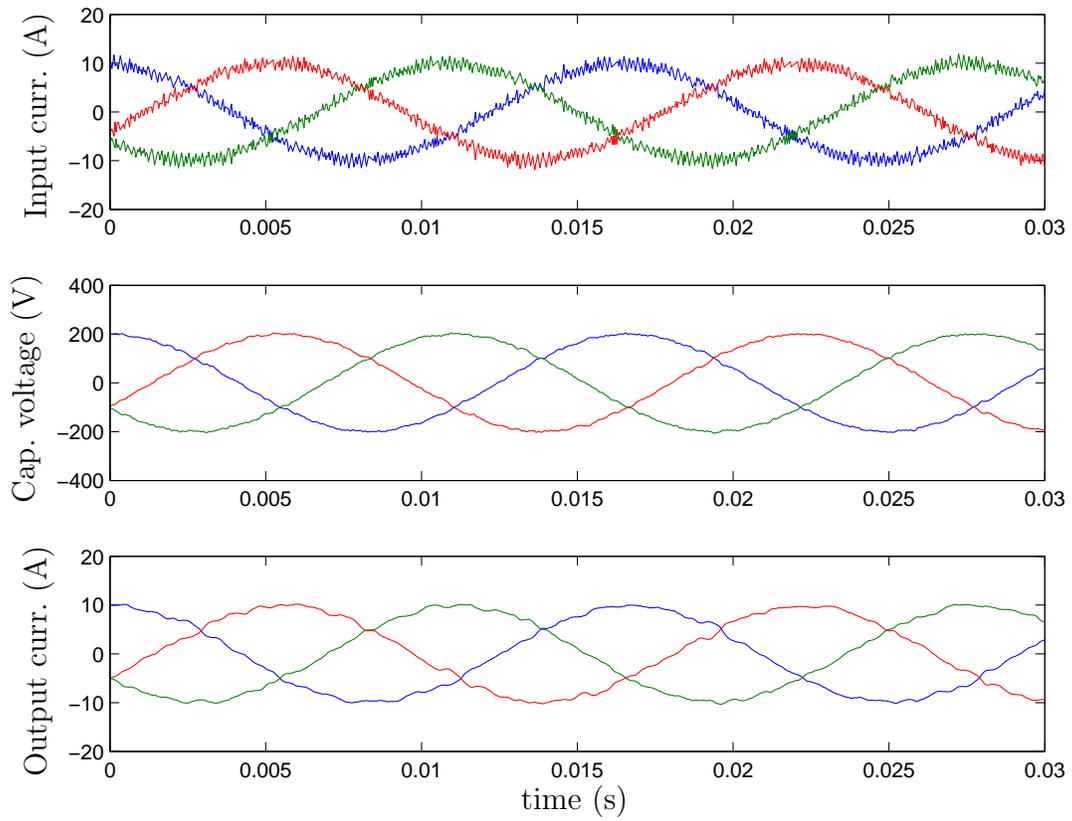}
\else
\includegraphics[width=.85\textwidth]{jl/adp_traj.eps}
\fi
\caption{
Closed-loop trajectories for the inverter example with the ADP policy,
for $\tau = 1$.
}
\label{f-inverter-traj-adp}
\end{figure}

\paragraph{Results.}
In figure~\ref{f-inverter-traj-adp}
we show the closed-loop response of the system
under the ADP controller with horizon length $T=1$.
The initial condition was taken to be in steady-state,
\ie, by choosing an initial state $x_0$,
computing the corresponding desired state 
$x_{\rm des} = Cx_0 + d$,
and starting the simulation from $x_{\rm des}$.

We compare the 1-step ahead ADP controller
to the FCS-MPC policy with $T = 10$ and $T = 30$,
solved to one percent accuracy.
We note that the system is unstable for $T < 5$,
and that a brute force solution of problem (\ref{e-multi-step-policy}) for $T \geq 5$
requires comparing $16807$ trajectories, 
which is not computationally feasible for real-time control.
In table~\ref{t-inverter}, 
we see the accumulated stage costs and switching costs over the simulation horizon,
for each of the the three control policies.
We see that the ADP controller performs comparably to the FCS-MPC policies,
both in terms of cumulative stage costs and switching costs.
In figure 
\ref{f-inverter-traj-adp},
we show the response using the A-MPC controller.

\begin{table}
\begin{center}
\begin{tabular}{lcc}
{\bf Control policy}             & {\bf State cost}   & {\bf Switching cost}   \\
\hline \hline                                                                 
ADP policy, $\tau = 1$           & 0.70               & 0.66        \\
FCS-MPC policy, $T = 5$          & 0.45               & 0.50        \\
FCS-MPC policy, $T = 10$         & 0.30               & 0.53        \\
\hline
\end{tabular}
\end{center}
\caption{
The average state cost and the average switching cost 
(\ie, the average values of $g(x_t)$ and $\ell(u_{t-1}, u_t)$
over the simulation),
for different control policies.
}
\label{t-inverter}
\end{table}



\section{Conclusion}
\label{s-conclucion}
We propose a method for approximately solving long-horizon FCS-MPC
problems that avoids extensive computation on the control processor.
Our method relies on estimating a quadratic approximate value function,
which we do by sampling.

We conclude by noting some other ways to obtain 
a quadratic approximate value function $\hat V$.
For example, if we minimize a quadratic cost function
over an approximate, linearized system,
a quadratic value function can be obtained by solving a Riccatti equation;
this method can be used instead of the function fitting method described in 
\S\ref{s-function-fitting}.
Another possible way involves solving linear matrix inequalities
corresponding to a relaxed Bellman equation;
for details, see \cite{rantzer2006relaxed} or \cite{wang2015approximate}.

\appendix

\clearpage
\bibliographystyle{alpha}
\bibliography{pow_elec_ctrl}

\clearpage
\section{Derivation of boost converter dynamics}
\label{s-boost-derivation}

%

Here we give the discrete dynamics for the boost converter example.
The discrete-time dynamics of the converter are approximated as:
\begin{align*}
f(x_t, u_t) = 
\begin{cases}
A^{\rm on,ccm} x_t + b^{\rm on,ccm} & u_t = 1,  \\
A^{\rm off,ccm} x_t + b^{\rm off,ccm} & u_t = 2, \; Cx_t \geq d \\
A^{\rm off,dcm} x_t & u_t = 2, \; Cx_t < d \\
\end{cases}
\end{align*}
We have
\begin{align*}
\begin{bmatrix}
A^{\rm on,ccm} & b^{\rm on,ccm}  \\
0 & 1
\end{bmatrix}
&=
\exp\left(
h \begin{bmatrix} -1/R_L L & -1/L & V_{\rm dc} \\ 1/C & -1/RC & 0 \\ 0 & 0 & 0 \end{bmatrix}
\right),
\\
\begin{bmatrix}
A^{\rm off,ccm} & b^{\rm on,ccm}  \\
0 & 1
\end{bmatrix}
&=
\exp\left(
h \begin{bmatrix} -1/R_L L & 0 & V_{\rm dc} \\ 0 & -1/RC & 0 \\ 0 & 0 & 0 \end{bmatrix}
\right),
\\
A^{\rm off,dcm} 
&=
\exp\left( h \begin{bmatrix} 0 & 0  \\ 0 & -1/RC  \end{bmatrix} \right),
\end{align*}
where $\exp$ denotes the matrix exponential.
The CCM/DCM threshold parameters $C$ and $d$ are chosen
to predict whether the inductor current drops below zero by the next time interval:
\begin{align*}
C = \begin{bmatrix} 1 & 0 \end{bmatrix} A^{\rm off,ccm},
\qquad
d = -\begin{bmatrix} 1 & 0 \end{bmatrix} b.
\end{align*}

\section{Derivation of inverter dynamics}
\label{s-inverter-derivation}
Here we show how to write the dynamics of the inverter in figure~\ref{f-inverter} 
in the form (\ref{e-sw-sys}).
The converter can be defined by the continuous-time differential algebraic equations
\begin{align*}
\dot x(t) &= A x(t) + B_{\rm in} v_{\rm in}(t)
+ B_{\rm load} v_{\rm load}(t) + Fv_{\rm float}
\\
0 &= F^T x
\end{align*}
where $x(t)\in\reals^9$ is the continuous system state,
consisting of all inductor currents and capacitor voltages in 
figure~\ref{f-inverter}:
\[
x(t) = (
i_1, 
i_2, 
i_3, 
v_1, 
v_2, 
v_3, 
i_4, 
i_5, 
i_6 
).
\]
Also, $v_{\rm in}(t)\in\reals^3$ is the vector of bridge voltages,
$v_{\rm load}(t)\in\reals^3$ is the vector of (sinusoidal) load voltages,
$v_{\rm float}(t) \in \reals^2$ is a vector of floating node voltages.
We have
\[
A = 
\begin{bmatrix} 
0 & 0 & 0 & -\frac1{L_1} & 0 & 0 & 0 & 0 & 0 \\
0 & 0 & 0 & 0 & -\frac1{L_1} & 0 & 0 & 0 & 0 \\
0 & 0 & 0 & 0 & 0 & -\frac1{L_1} & 0 & 0 & 0 \\
\frac1C & 0 & 0 & 0 & 0 & 0 & -\frac1C & 0 & 0 \\
0 & \frac1C & 0 & 0 & 0 & 0 & 0 & -\frac1C & 0 \\
0 & 0 & \frac1C & 0 & 0 & 0 & 0 & 0 & -\frac1C \\
0 & 0 & 0 & -\frac1{L_2} & 0 & 0 & 0 & 0 & 0 \\
0 & 0 & 0 & 0 & -\frac1{L_2} & 0 & 0 & 0 & 0 \\
0 & 0 & 0 & 0 & 0 & -\frac1{L_2} & 0 & 0 & 0 \\
\end{bmatrix},
\qquad
B_{\rm in} = 
\begin{bmatrix} 
\frac1{L_1} & 0 & 0 \\
0 & \frac1{L_1} & 0 \\
0 & 0 & \frac1{L_1} \\
0 & 0 & 0 \\
0 & 0 & 0 \\
0 & 0 & 0 \\
0 & 0 & 0 \\
0 & 0 & 0 \\
0 & 0 & 0 \\
\end{bmatrix},
\]
\[
B_{\rm load} = 
\begin{bmatrix} 
0 & 0 & 0 \\
0 & 0 & 0 \\
0 & 0 & 0 \\
0 & 0 & 0 \\
0 & 0 & 0 \\
0 & 0 & 0 \\
-\frac1{L_2} & 0 & 0 \\
0 & -\frac1{L_2} & 0 \\
0 & 0 & -\frac1{L_2} \\
\end{bmatrix},
\quad
F =
\begin{bmatrix} 
-\frac1{L_1} & -\frac1{L_1} \\
-\frac1{L_1} & -\frac1{L_1} \\
-\frac1{L_1} & -\frac1{L_1} \\
0 & 0 \\
0 & 0 \\
0 & 0 \\
0 & -\frac1{L_2} \\
0 & -\frac1{L_2} \\
0 & -\frac1{L_2} \\
\end{bmatrix}.
\]
We can solve these differential algebraic equations to obtain:
\[
\dot x(t) = A x(t) + B_{\rm in} v_{\rm in}(t) + B_{\rm load} v_{\rm load}(t),
\]
where 
$A = A - M A$, 
$B_{\rm in} = B_{\rm in} -  M  B_{\rm in}$,
$B_{\rm out} = B_{\rm out} - M   B_{\rm out}$,
and $M = F(F^T F)^{-1} F^T$.

We now show how to represent the time varying, uncontrolled vector
$v_{\rm load}$ in terms of $\sin \omega t$ and $\cos \omega t$.
We have
\[
v_{\rm load} 
= 
V_{\rm load}
\begin{bmatrix}
\sin (\omega t) \\ 
\sin (\omega t + \frac{2\pi}{3}) \\ 
\sin (\omega t + \frac{4\pi}{3}) \\ 
\end{bmatrix}
= 
V_{\rm load}
\begin{bmatrix}
1 & 0 \\ 
\sin\frac{2\pi}{3} & \cos\frac{2\pi}{3} \\
\sin\frac{4\pi}{3} & \cos\frac{4\pi}{3} \\
\end{bmatrix}
\begin{bmatrix}
\sin \omega t \\ 
\cos \omega t \\ 
\end{bmatrix}.
\]
By incorporating $\sin \omega t$ and $\cos \omega t$ as state variables,
we can then express the system in a time-invariant form:
\begin{align*}
\frac{d}{dt} \begin{bmatrix} x(t) \\ \sin \omega t \\ \cos\omega t \end{bmatrix}
=
\left[ \begin{array}{c|cc}
A & \multicolumn{2}{c}{B_{\rm load}} \\
\hline
\multirow{2}{*}{0} & 0 & -\omega \\
& \omega & 0
\end{array} \right]
\begin{bmatrix} x(t) \\ \sin \omega t \\ \cos\omega t \end{bmatrix}
+
B_{\rm in} v_{\rm in}(t).
\end{align*}
Because the input voltage is constant over each time epoch,
we can exactly discretize this differential equation using a matrix exponential:
\[
\begin{bmatrix}
A & B_{\rm load} \\
0 & I
\end{bmatrix}
=
\exp\left(
\Delta t
\begin{bmatrix}
A & B_{\rm load} \\
0 & 0
\end{bmatrix}
\right)
\]
where $\Delta t$ is the discretization timestep.

Note that in each time epoch,
$v_{\rm in}(t)$ can take one of seven values:
\[
\begin{bmatrix} 0 \\ 0 \\ 0 \end{bmatrix},
\quad
\begin{bmatrix} V_{\rm dc} \\ 0 \\ 0 \end{bmatrix},
\quad
\begin{bmatrix} 0 \\ V_{\rm dc} \\ 0 \end{bmatrix},
\quad
\begin{bmatrix} 0 \\ 0 \\ V_{\rm dc} \end{bmatrix},
\quad
\begin{bmatrix} V_{\rm dc} \\ V_{\rm dc} \\ 0 \end{bmatrix},
\quad
\begin{bmatrix} V_{\rm dc} \\ 0 \\ V_{\rm dc} \end{bmatrix},
\quad
\begin{bmatrix} 0 \\ V_{\rm dc} \\ V_{\rm dc} \end{bmatrix}.
\]
Premultiplying these vectors by $B_{\rm load}$ yields
the vectors $b^{u_t}$ in equation 
(\ref{e-lin-sys-sw-input}).

%
%
%

\end{document}